\documentclass[11pt,twoside]{article}
\usepackage{asp2004}
\usepackage{epsf}
\usepackage{psfig}
\usepackage{lscape}
\begin{document}

\title{Cyclic Period Changes in Close Binaries:\\
A Light Travel Time Effect or a\\ 
 Symptom of Magnetic Activity?}

\author{R.~T. Zavala}
\affil{United States Naval Observatory, Flagstaff Station, P.O. Box
1149, Flagstaff, AZ USA 86004; bzavala@nofs.navy.mil}

\begin{abstract}
Years to decade-long cyclic period changes have been observed in many 
classes of close binaries. A common explanation invoked for these cyclic 
period changes is the gravitational influence of an unseen third body upon 
the close binary pair.  The effects of an unseen third body must be strictly 
periodic, and not all of these observed period variations match this 
prediction. Douglas Hall noted almost 20 years ago that Algol binaries with 
cyclic period changes always had a convective, rapidly rotating secondary star. 
This secondary star has the seeds for a powerful magnetic dynamo, and the 
suggestion was made that magnetic activity was the cause of the observed 
cyclic period changes. 

In this paper I will review the observational evidence for a magnetic cycle 
driving the observed cyclic period changes. I will also discuss the theoretical 
framework which now exists to explain how the magnetic cycle can alter the 
orbital period of a close binary. I will close by discussing an ongoing 
observational effort to test the predictions of the magnetic activity-cyclic 
period change connection.   
\end{abstract}

\section{Introduction}
The light travel time effect (LTTE) manifests itself as a periodic  
oscillation in a plot of observed minus calculated ($O-C$) times 
of phenomena in variable stars. Such an oscillation in the $O-C$ 
plot may reveal the presence of an otherwise unseen companion object(s).
A particularly pathological example of the effects of a third body is 
the cessation of eclipses in SS Lac \citep{ts00,ek01}.
Cyclic orbital period changes revealed in $O-C$ plots for eclipsing 
binaries are sometimes interpreted as the manifestation of an unseen 
companion on the close binary pair. Long-term cyclic period changes are 
a fairly common phenomenon in close binary systems. Algol \citep{hall89}, 
RS CVn \citep{hk80}, and W UMa \citep{hob94} binaries all exhibit cyclic 
period changes on decade long timescales. While some of these close 
binaries with cyclic $O-C$ residual plots do have distant companions the 
LTTE is not always a good explanation. This was noted as early as 1939 
by \citet{dw39}. 

In \S2 I will attempt to 
illustrate how the LTTE fails to consistently explain $O-C$ data for Algol, 
RS CVn and W UMa binaries. In \S3 I present some non-LTTE driven mechanisms  
which fail to explain cyclic period changes.
In \S4 I will present review the observational and theoretical support 
for a magnetic cycle driven model for cyclic period changes. Suggestions for future 
work appear in \S5. 
 
\section{Problems for the LTTE Hypothesis}
 
A key prediction of the LTTE of an unseen companion is that the 
period variation must be strictly sinusoidal \citep{kop78}. Thus, the year to 
decades long cyclic orbital period changes in close binaries must be a regular 
sinusoidal function. Given enough patience on the part of the observer, 
or a thorough literature search, one can examine the period changes in 
close binaries for evidence of a strictly sinusoidal residual in the $O-C$ plots.
The prototypical Algol $\beta$ Persei provides the longest time baseline 
I am aware of for examining these cyclic period variations. Algol displays a 
32 year period in its $O-C$ data which \citet{fhh} interpreted as apsidal motion. 
The close binary pair in Algol, Algol AB, has a circular orbit \citep{sod80} 
and is thus not likely to show apsidal motion. Additionally, the primary and 
secondary eclipses of Algol AB do not shift in opposite directions as expected 
if apsidal motion is the culprit. The lack of a regular sinusoidal pattern 
in the $O-C$ data for Algol AB \citep{sod80} makes an LTTE explanation 
less than satisfactory.

It is possible to infer the parameters of an unseen companion responsible 
for a supposed LTTE variation in the $O-C$ residuals of an eclipsing binary.   
\citet{vb86} used the eclipse timing residuals for 8 RS CVn systems and 
determined that only two systems had plausible third bodies. In some systems he 
found the $O-C$ data implied a neutron star or black hole companion in order to be 
consistent with the lack of detection of the third body. Although black holes 
and neutron stars certainly exist the space density of RS CVn systems presents 
difficulties when one considers the number of massive star progenitors required
\citep{vb86}. 

\citet{bheg96} carried out a Fourier 
analysis of 18 close binaries of various classes, and found good agreement
to the third body hypothesis for four of the systems. Borkovits and 
Heged$\rm\ddot{u}$s were careful to check that the resulting solutions were 
physically realistic and compatible with the lack of detection of the third 
body. Even so, we see a trend emerging that the LTTE hypothesis does not explain 
all the cyclic period changes observed in close binaries. As different classes 
of close binaries exhibit the same cyclic period changes a common explanation 
in the spirit of Occam's Razor would be most satisfying. The Light Travel Time 
Effect fails as a common explanation and researchers eventually turned 
to other explanations.

\section{Non-LTTE Explanations for Cyclic Period Changes} 
Mass transfer between stars in an interacting binary can cause period 
changes, but does not seem to explain cyclic period variations. 
In the case of mass transfer which conserves angular momentum and mass 
\citet{kwee58} showed that mass transfer can only increase {\it or} decrease 
the orbital period, and cannot cause cyclic period variations. \citet{bh73} 
suggested sudden bursts of mass transfer which temporarily stored 
angular momentum in the rotation of the hot mass-gaining star in Algol systems 
as a possible cause of cyclic period variations. The sudden mass transfer event 
rapidly decreases the orbital period, which then slowly increases as 
tidal forces return the angular momentum to the binary system.    
Subsequent work on the Algol U Cep \citep{ols81} failed to demonstrate the 
gradually changing period required by \citeauthor{bh73}. Non-conservative 
mass transfer could cause period changes, but not of the cyclic variety 
seen in the $O-C$ plots \citep{th91}. By 1991 the mass transfer ideas had begun 
to be supplanted by a magnetic cycle driven model. 

\section{Magnetic Cycles and Cyclic Period Changes}
In his paper which provided a precise definition of the RS CVn class 
\citet{hall76} noted that both Algols and RS CVn binaries exhibited cyclic 
period changes of a similar magnitude, but he suspected at this point that 
the causes were different. In 1976 the period change mechanism for Algols 
was connected with mass transfer. RS CVn binaries, which are detached and thus 
no Roche lobe overflow enables mass transfer, were hypothesized to undergo 
smaller levels of mass loss perhaps related to winds.  

A significant development in our understanding of the cyclic period 
changes was made by Douglas Hall (1989). In his paper entitled ``The Relation 
between RS CVn and Algol'' Hall argued that magnetic activity cycles 
were the cause of the observed cyclic period variations.

\begin{figure}[!ht]
\plotone{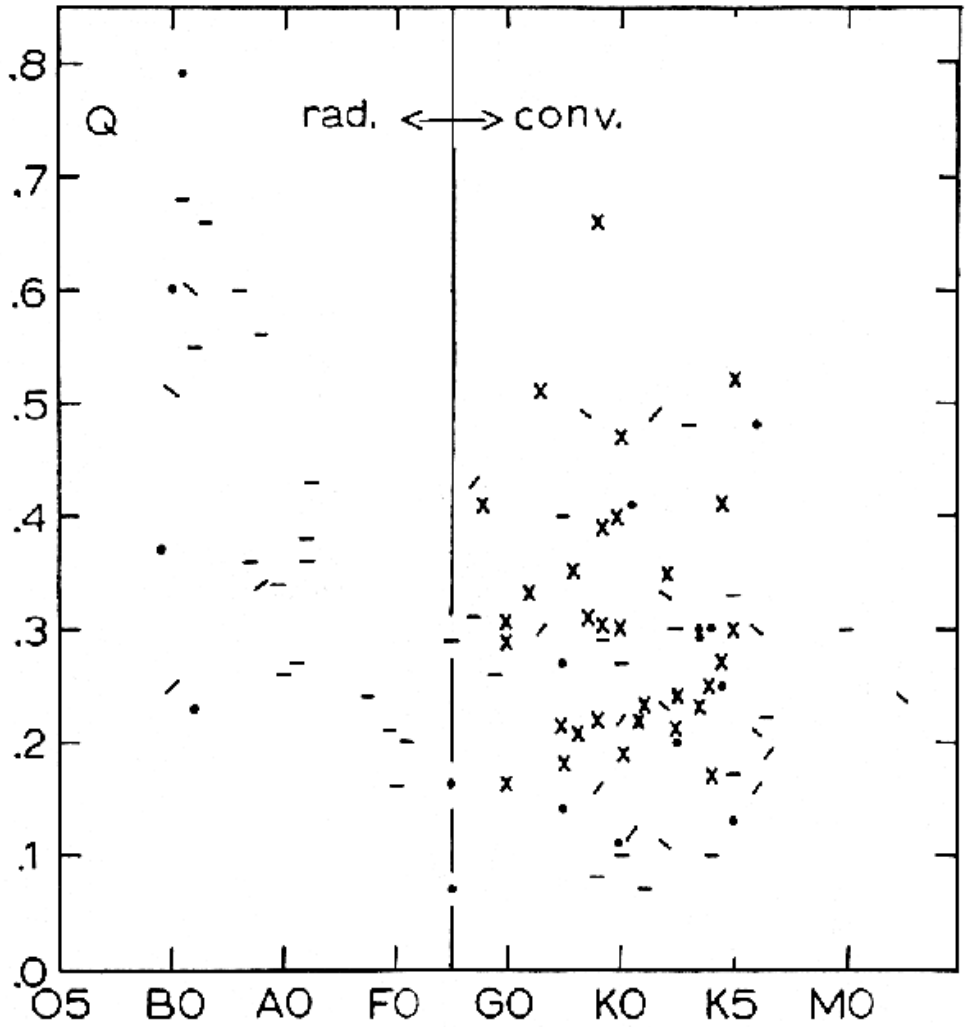}
\caption{Reproduction of Figure 4 from \citet{hall89}. The character of 
orbital period changes in Algol binaries is shown in this plot of mass 
ratio Q to secondary star spectral type. A constant period is indicated 
with a $-$, increasing period $/$, decreasing period $\setminus$, and 
alternating period increases and decreases by X. A {$\bullet$} indicates 
systems for which no conclusion could be drawn. The vertical line in the 
middle of the plot separates stars with radiative outer atmospheres from 
those with convective outer atmospheres.}
\label{hall}
\end{figure}

The key piece of evidence in support of magnetic activity as a cause 
of cyclic period changes is shown in Fig.~\ref{hall}. Fig.~\ref{hall} 
shows the mass ratio Q for over 100 Algol binaries versus the spectral type of the 
secondary star in the binary. Hall shows the types of period changes observed in these 
Algols by using different symbols which are explained in the caption. The significant 
finding of Hall was that all (31 out of 31) cases of alternating period increases 
{\it and} decreases occur in systems which have secondary stars later than spectral 
type F5. These secondary stars have convective outer atmospheres, and combined 
with rapid stellar rotation from spin-orbit coupling the ingredients for a stellar 
magnetic dynamo are present \citep{par79}. \citet{hall89} proposed a magnetic activity 
cycle driven by these stellar dynamos to explain cyclic period changes in both 
RS CVn and Algol binaries, although the details of the mechanism were not specified.

\begin{figure}[!ht]
\plotone{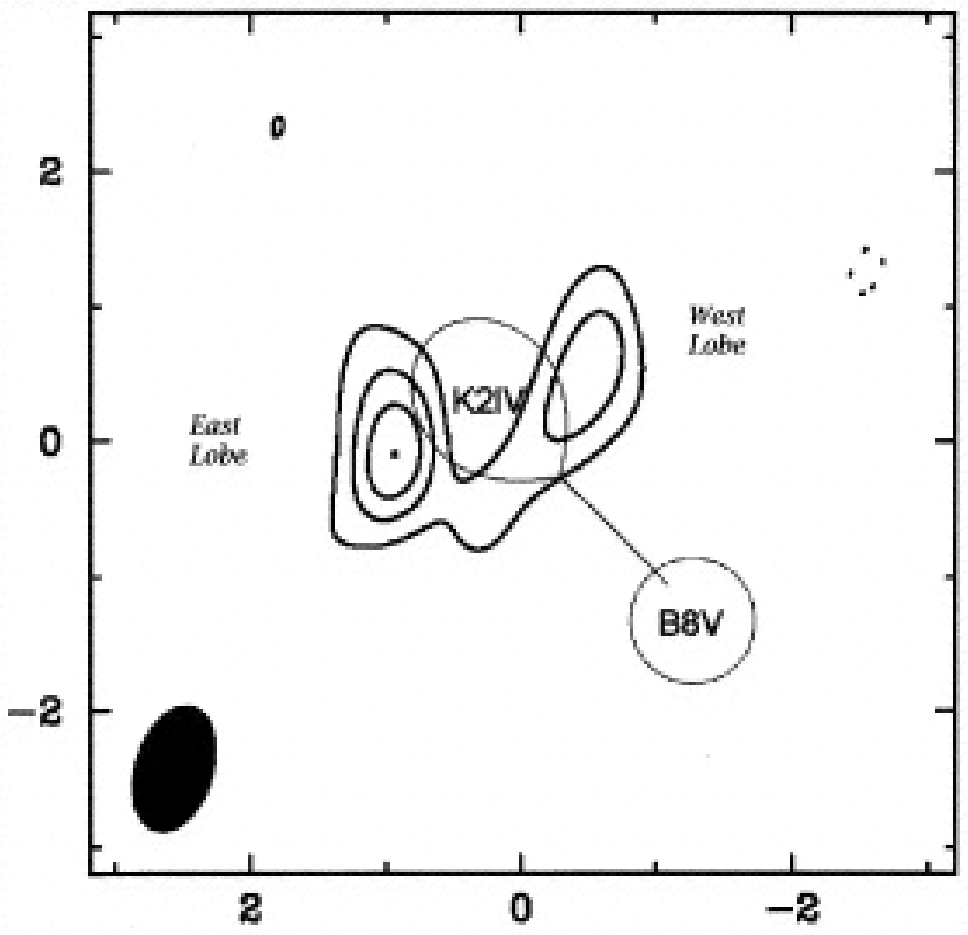}
\caption{Stokes I VLBA radio image of Algol at 8.4 GHz from \citet{rlm}.
The shaded oval in the lower left is the FWHM of the synthesized beam 
of the VLBA. X and Y axes are $\alpha$ and $\delta$ relative to the image 
center. Contours are 0.8, 1.2, 1.6 and 2.0 mJy beam$^{-1}$.}
\label{robert}
\end{figure}

If magnetic activity is responsible for the cyclic period changes in Algols than 
Algols should exhibit observational characteristics similar to the magnetically active 
RS CVn binaries. Infrared light curves of the secondary star of Algol show variations 
similar to that seen in the spotted RS CVn stars \citep{mr90}. The Very Long Baseline 
Array image of \citet{rlm} shows lobes of emission above the poles of the secondary 
star of Algol which are interpreted as gyrosynchrotron radiation (Fig.~\ref{robert}). 

\begin{figure}[!ht]
\plotone{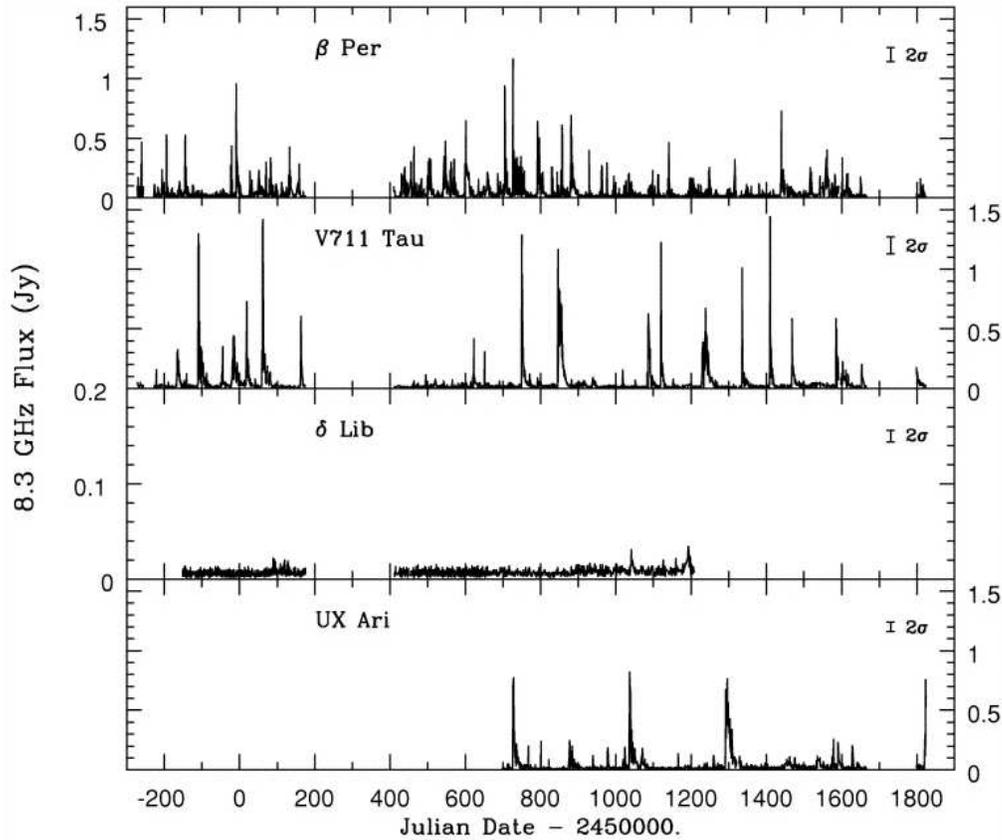}
\caption{8.4 GHz radio lightcurves for 2 Algol (Algol and $\delta$ Lib) and 2 RS CVn stars 
(V711 Tau and UX Ari) from \citet{rwgr}. The lightcurves were obtained using the 
two element Green Bank Interferometer from 1995$-$2000.}
\label{rlc}
\end{figure}

\citet{mrga93} 
presented observational evidence for RS CVn-like magnetic activity in Algols from 
radio, optical and x-ray indicators. The radio luminosity functions of RS CVn and 
Algol systems appears similar \citep{utc} but small number statistics may leave 
this result suspect. A long-term radio monitoring survey 
showed that both Algol and RS CVn stars have similar levels of radio flare activity 
\citep{rwgr}. The radio lightcurves for 2 Algol and 2 RS CVn stars from \citeauthor{rwgr} 
is reproduced in Fig.~\ref{rlc}. Algol itself had slightly more frequent, though 
slightly weaker, radio flares as compared to the RS CVn star V711 Tau. 

A theoretical basis for magnetically induced cyclic period variations was introduced 
by \citet{app87}. They proposed a model in which deformations of the active 
star away from hydrostatic equilibrium changed the gravitational quadrupole moment
of the secondary star. As a star went through its magnetic activity cycle the 
gravitational quadrupole moment was cyclically changed and the period changes reflected 
the changing quadrupole moment. This model soon required modification when 
\citet{mp90} showed that deviations away from hydrostatic equilibrium were 
ruled out on energetic grounds. 

The energetic problem was solved when \citet{app92} proposed a model which 
connected variations of the gravitational quadrupole moment with the distribution 
of angular momentum in the active star. This required some of the energy from 
active star's luminosity to be diverted into differential rotation. 
Changes in differential rotation then changed the oblateness of the active star 
and thus the quadrupole moment changed. This diversion of energy from 
luminosity into differential rotation meant that a cyclic variation in 
the luminosity of the active star should also occur. The active star 
transitions between states of hydrostatic equilibrium and therefore radial 
pulsations are not required. Only differential rotation changes, not the 
mean radius. A color change is created as the luminosity changes at constant radius, 
so as luminosity increases the star should become bluer. In summary \citet{app92} 
makes three predictions for his magnetic activity model: the light curve and $O-C$ curve 
should both have cycles of the same length, extrema in brightness should 
coincide with extrema in the $O-C$ diagram, and the star should be bluest when it 
is brightest. This model was placed on a firmer theoretical basis by the work of 
\citet{lrr98} and \citet{rud02}. Models for the magnetic dynamos responsible for the 
cyclic period changes are presented in these two works, and estimates of the sub-surface 
magnetic field strengths of a few kilogauss are made. This theoretical framework than 
provides a method for probing the internal magnetic structure of stars in close binaries.

Applegate's three predictions were observed in the RS CVn binary CG Cyg by  
\citet{hall91}. Subsequently \citet{ahi04} obtained new photometric measurements 
of CG Cyg and they claim that two periodicities exist in CG Cyg. A period of 
52 years in the $O-C$ curve is interpreted as evidence for a light travel time effect 
from a third body of 0.9 M$_{\odot}$. \citeauthor{ahi04} concur that there is 
a variation of the average brightness of CG Cyg, but assign to this brightness 
variation a period of 22.5 years. Thus, CG Cyg may present an interesting case where
both the light travel time effect and a magnetic cycle are present. Verification of 
a magnetic cycle induced period change by means of a varying mean brightness and color 
is complicated by the long timescale of the variation. Over the decades detector 
technology changes radically, and different comparison stars complicate the analysis 
\citep{ahi04}. 

The Algol binary WW Cyg also apparently satisfies Applegate's three 
predictions, but the luminosity and color variations are undersampled 
in time \citep{zav02,zav04}. In the case of WW Cyg I and my collaborators could 
not obtain a satisfactory fit to a third body orbit, or a third and fourth body orbit 
\citep{zav02}. Our preliminary analysis of the color and luminosity variation of 
WW Cyg at primary eclipse minima did agree with the predictions of the magnetic activity 
hypothesis. In Table 1 we present a
more quantitative analysis of the luminosity and color variations of
WW Cyg at primary eclipse minimum. The V magnitudes in Table 1 were produced
by averaging data points within 0.0066 days of eclipse midpoint. This is just
over 3$\sigma$ of the worst time of minima estimate in \citet{zav02}.
This yields 35 data points in V obtained from 1997 to 1999, and 4 data points
for \citet{haw72}. \citet{haw72} used $UBV$ filters, and \citet{zav02} used
$VRI$, thus direct comparisons in color index $B-V$ are not possible.
The de-reddened $B-V$ of 0.82 of \citet{haw72} corresponds to a $V-R$ color
of 0.42 \citep{allen}, or a K0V spectral type. We estimated the error in
Hall \& Wawrukiewicz's $V-R$ from the errors presented in their paper.
Our estimate of the color of WW Cyg at primary minimum 25 years
after \citet{haw72} is significantly redder, and corresponds to a main
sequence spectral type of K2V. This agrees with the prediction of
\citeauthor{app92} that the active star should be reddest when it is faintest,
and supports a magnetic cycle driven origin for the period changes in WW Cyg.

\begin{table}[!ht]
  \caption{Magnitude and Color Changes of WW Cyg  }
  \label{tab:zav1}
  \smallskip
  \begin{center}
  {\small
  \begin{tabular}{ccc}
  \tableline
  \noalign{\smallskip} 
    Date & V & V$-$R \\
  \noalign{\smallskip}
  \tableline
  \noalign{\smallskip}
    1972 & $13.26\pm0.01$ & $0.42\pm0.02$\\
    1997$-$1999 & $13.381\pm0.003$ & $0.48\pm0.01$\\
  \noalign{\smallskip}
  \tableline
  \end{tabular}
  }
\end{center}
\end{table}

\section{Future Work}

Photometric observations of 11 Algol, W UMa, and RS CVn systems are
currently underway to test the magnetic activity hypothesis. These systems
all show cyclic period variations in their $O-C$ diagrams, and we are
obtaining multi-color photometry to examine the correlation between luminosity
and color variations as predicted by Applegate. The observations are underway 
and use the 40 inch (1 meter) reflector at the U.S. Naval Observatory, 
Flagstaff Station. 
                                                                                           
A further interesting case is that of $\delta$ Librae, an Algol binary with
a third body suggested by \citet{worek}. This third body is predicted to have
a mass of $\approx$ 1$M_\odot$ and a period of 2.76 years. The radio lightcurve 
of $\delta$ Lib looks relatively quiet compared to Algol (Fig.~\ref{rlc}), but 
$\delta$ Lib may simply have been in a magnetically quiet stage during the observations. 
$\delta$ Lib is more distant than Algol, but if Algol were at the same distance it 
would still have a radio flux of 100 mJy, so $\delta$ Lib is certainly radio quiet 
compared to Algol. The period of 2.76 years proposed by \citeauthor{worek} 
is shorter by a factor of 10 
compared to the cyclic period change timescales for Algols. Thus, $\delta$ Lib 
may present an interesting case of a close binary which should exhibit magnetic 
cycles, but fails to do so. The predicted amplitude of the barycenter motion is 
approximately 5 mas, or 1.8 mas yr$^{-1}$. The Navy Prototype Optical Interferometer
\citep[NPOI,][]{npoi} will eventually have 437 m baselines and a
resolution of 0.3 mas, and will thus be able to resolve the motion
expected by a third body. The superb resolution of optical long baseline
interferometry presents a direct method for examining the source of the
cyclic period changes in close binaries. The small amplitude of the motion 
about the barycenter can be measured and the LTTE hypothesis put to a direct test. 
When combined with the brightness and color variation predictions optical interferometry 
can provide a convincing test of the validity of magnetically driven cyclic period 
changes in close binaries.  
 
\acknowledgements{I thank the organizers of the conference and the United 
States Navy for support which made my attendance possible. Brenda Corbin and Gregory 
Shelton of the U. S. Naval Observatory Library were very helpful in providing 
references for this paper. This research has made use of the SIMBAD database 
operated at CDS, Strasbourg, France and the NASA Astrophysics Data System.}

\end{document}